\documentstyle[epsf]{article}

\newcommand{\jinft}{\mbox{$
\,{}_{{}_{N\rightarrow\infty}}
\!\!\!\!\!\!\!\!\!\!\!\!\longrightarrow\,$}}
\newcommand{\lb}{\mbox{\bf (}}
\newcommand{\rb}{\mbox{\bf )}}
\newcommand{\mb}{\mbox{$|\!\!|$}}

\title{Classical Resonances and Quantum Scarring}

 \author{Christopher Manderfeld
\\Fachbereich Physik, Universit\"at Essen, 45117 Essen, Germany}

\begin{document}

\maketitle
\begin{abstract}
We study the correspondence between phase-space localization 
of quantum (quasi-)energy eigenstates 
and classical correlation decay, 
given by Ruelle-Pollicott resonances of the Frobenius-Perron operator. 
It will be shown that 
scarred \mbox{(quasi-)}energy eigenstates are correlated: 
Pairs of eigenstates strongly overlap in phase space 
(scar in same phase-space regions) 
if the difference of their 
eigenenergies is close to the phase of a leading classical resonance. 
Phase-space localization of quantum states will be measured by $L^2$ 
norms of their Husimi functions. 
\end{abstract}
\vspace{1 cm}

\section{Introduction}
Schnirelman's theorem \cite{schnir} as well as Berry's physical reasoning 
predict that quantum energy eigenfunctions 
(Wigner or Husimi representation \cite{wigner,husimi})
of systems whose classical counterpart is chaotic are uniformly distributed on 
the energy shell. 
Heller, however, has shown that there exist quantum 
eigenfunctions which are strongly localized (scarred) on hyperbolic periodic 
orbits \cite{heller1}. 
At present, the general opinion is that scars are exceptional, 
while the majority consits of uniformly (phase-space) distributed 
eigenstates \cite{voros}. 
On the other hand, it is well known that there also exist weakly localized 
eigenfunctions, which continuously fill the ``gap'' between uniformly distributed 
and strongly localized eigenfunctions. 
Conveniently, instead of single exceptions we here consider localization 
properties of the whole set of eigenfunctions of the system; 
restricting our studies on finite dimensional Hilbert spaces. 
To that purpose, we introduce squared $L^2$ norms of  Husimi eigenfunctions 
as a measure for phase-space localization \cite{gnutz}. 
As it will be shown in the sequel this measure proves amenable to 
semiclassical considerations. 

The classical dynamics of chaotic systems can be described by the time evolution 
of phase-space density functions. The corresponding propagator is the 
Frobenius-Perron operator \cite{hasegawa,lasota,gaspard}, 
where the poles of its resolvent are called 
Ruelle-Pollicott resonances 
which coincide with decay rates of classical correlation functions 
\cite{pollicott,ruelle1,ruelle2,ruelle3}.  
The quantum-classical correspondence, in particular the influence of Ruelle-Pollicott resonances 
on the quantum energy spectrum, is still a point of interest \cite{altsh1,altsh2,altsh3,altsh4}.
We show that phase-space overlaps of energy eigenstates turn out Lorentz distributed with 
respect to the differences of their eigenenergies. 
The Lorentzians are determined by Ruelle-Pollicott resonances. 
In other words, the probability that two eigenstates strongly overlap becomes large if the difference 
of their eigenenergies coincides with the position where the Lorentzian is peaked. 
On the other hand, if a pair of eigenstates strongly overlaps 
(much more than random-matrix theory predicts) each of them must be localized, i.e. scarred, 
in the same phase-space regions.

We here consider systems with a compact 
two-dimensional phase space, in particular the unit sphere, whereby 
the Hilbert-space dimension of the quantum counterpart becomes finite. 
Periodically driving destroyes integrability in general. 
Moreover, a stroboscopic description 
leads to a Hamilton map or a Floquet operator in the classical or 
quantum case, respectively. A well known representative of 
such a dynamics is the kicked top \cite{scharf,braun,kus2}.

\section{Kicked Top}
\label{kickedtop}
The dynamics of the kicked top is described by a stroboscopic 
map of an angular momentum vector whose length is conserved, 
For such a dynamics the phase spaces is the unit sphere. 
The classical time evolution is usually described in the ``Hamilton picture'', 
whereby the stroboscopic consideration leads to a Hamilton map which 
describes a trajectory after each period, 
\begin{equation}
(q',p')={\cal M}(q,p)
\,.
\end{equation}
Here the primes denote the final position $q$ and momentum $p$ coordinates. 
On the sphere the canonical phase-space coordinates are given by 
the azimuthal and polar angle as $q=\varphi$ and $p=\cos\theta$. 

The quantum time evolution in the Schr{\"o}dinger picture 
is generated by a Floquet operator $F$ 
which is built by the components of an 
angular momentum operator ${\bf J}$; 
we choose it as a sequence of rotations about the $y$ and $z$-axis
followed by a nonlinear torsion about the $z$-axis, 
\begin{eqnarray}
F&=&T_z(\tau)R_z(\alpha)R_y(\beta)\,,\nonumber\\
&&T_z(\tau)={\rm e}^{-{\rm i}\frac{\tau}{N}J_z^2}\,,\nonumber\\
&&R_z(\alpha)={\rm e}^{-{\rm i}\alpha J_z}\,,\nonumber\\
&&R_y(\beta)={\rm e}^{-{\rm i}\beta J_y}\,,
\end{eqnarray} 
where $\tau$ is called the torsion strength and $\alpha$ and $\beta$ are 
rotation angles. 
The dimension of the quantum Hilbert space is $N=2j+1$, where $j$ is 
the quantum angular momentum formally replacing the inverse of
Planck's constant, $\hbar^{-1}$ ($\hbar=1$ in this article).  
Since $F$ is unitary, it has $N$ orthogonal eigenstates with 
unimodular eigenvalues characterized by eigenphases (quasi-eigenenergies) 
as $F^n|\phi_i\rangle=\exp(-{\rm i}n\phi_i)|\phi_i\rangle$. 

The classical counterpart has in general a mixed phase space. 
By the choice of the parameters as $\alpha=\beta=1$ and $\tau=10$ the 
dynamics becomes strongly chaotic. For $j=200$ which we use for numerical results 
stable island are not resolved by the Planck cell of size $\frac{4\pi}{N}$, whereby 
the dynamics looks, from a quantum point of view, effectively hyperbolic. 

In order to compare the results of the kicked top with those of random-matrix theory (RMT) 
we here discuss the symmetries of the system. 
The dynamics proves invariant under nonconventional time reversal. In terms of 
random-matrix theory the Floquet matrix belongs to the circular orthogonal ensemble (COE), 
where the coefficients of the eigenvectors can be chosen real in a suitable basis \cite{dyson,mehta}. 
We here expand the Floquet operator in the basis of eigenstates of the $z$-component of the 
angular momentum operator, $J_z|jm\rangle=m|jm\rangle$. 
By a unitary transformation given by a simple rotation, $F'=\tilde{R}F\tilde{R}^\dagger$, the Floquet matrix 
becomes symmetric, ${F'}^T=F'$, where the eigenvectors become real. 
Here $T$ denotes transposition. This is an important property, since 
the $L^2$ norm of a Husimi function which we will use 
as a measure for phase-space localization 
is invariant under rotations. Therefore, eigenvectors of the kicked top must be compared to real 
random vectors. 
The rotation is of form 
 $\tilde{R}=R_z(\frac{{}_\beta}{{}^2})R_y(-\frac{{}_\pi}{{}^2})R_z(-\frac{{}_\pi}{{}^2})$, 
whereby the transformed Floquet matrix becomes
\begin{equation}
F'=R_z(\frac{{}_\beta}{{}^2})R_y(-\frac{{}_\pi}{{}^2})R_z(-\frac{{}_\pi}{{}^2})
T_z(\tau)R_z(\alpha)R_y(\beta)R_z(\frac{{}_\pi}{{}^2})R_y(\frac{{}_\pi}{{}^2})
R_z(-\frac{{}_\beta}{{}^2})
\,.
\end{equation}
After commutation of the rotation $R_z(-\frac{{}_\pi}{{}^2})$ with the torsion the product 
\begin{equation}
R_z(-\frac{{}_\pi}{{}^2})R_y(\beta)R_z(\frac{{}_\pi}{{}^2})=
R_y(\frac{{}_\pi}{{}^2})R_z(\beta)R_y(-\frac{{}_\pi}{{}^2})
\end{equation}
is a rotation about the $x$-axis. Using the latter relation the transformed 
Floquet matrix finally becomes 
\begin{equation}
F'=\Big(R_y(\frac{{}_\pi}{{}^2})R_z(\frac{{}_\beta}{{}^2})\Big)^T
T_z(\tau)R_z(\alpha)
R_y(\frac{{}_\pi}{{}^2})R_z(\frac{{}_\beta}{{}^2})
\,,
\end{equation}
which is obviously symmetric.

\section{Frobenius-Perron Operator}
Another way to describe classical time evolution is the ``Liouville picture'', 
as the propagation of density in phase space. 
The corresponding propagator, the Frobenius-Perron operator ${\cal P}$, 
is defined through the Hamilton map as
\begin{equation}
f_n(q,p)={\cal P}f_0(q,p)=f_0({\cal M}^{-n}(q,p))
=\int{\rm d}q'\,{\rm d}p'\,f_0(q',p')\delta^2((q',p')-{\cal M}^{-n}(q,p))
\,,
\label{fpoper}
\end{equation}
where $f(q,p)$ denotes an arbitrary phase-space density function. 
Note that the Hamilton map is invertible and area preserving. 
An expectation value of a classical observable is given by the 
phase-space integral 
\begin{equation}
\left\langle A\right\rangle=
\int{\rm d}q\,{\rm d}p\,A(q,p)f(q,p)
\equiv\lb A\mb f\rb
\,,
\label{average}
\end{equation}
where we have introduced the Dirac notation. 
To avoid confusion with 
quantum wave functions we here use round brackets. Note that this 
notation is generally to read as a linear functional, where the 
density function $f$ belongs to the Banach space $L^1$ and the 
observable $A$ to the dual space $L^\infty$. Furthermore, we 
suppose that both functions are real, otherwise $A$ is to complex 
conjugate in the integral notation (\ref{average}). 

For classically chaotic systems (we assume purely hyperbolic dynamics) 
correlations of observables decay exponentially in time. 
Due to ergodicity the time correlation can be written by a phase-space integral,
\begin{equation}
C_{AB}(n)=\lb A(n) B(0)\mb\rho_i\rb-\lb A\mb \rho_i\rb\lb B\mb \rho_i\rb
\,,
\end{equation}
where the time dependence of the observables must be read as 
$A(n)=A(q(n),p(n))$. 
Here $\rho_i$ denotes the stationary (invariant) density
with ${\cal P}\rho_i=\rho_i$, i.e. the constant on 
the sphere. The associated stationary eigenvalue is $1$ and ensures that 
no probability gets lost, i.e. it preserves the $L^1$ norm of a density function. 
We may replace the observable $B$ by an initial density function $f$ and further we 
assume that $\lb A\mb \rho_i\rb=0$, then the correlation function can be written 
in terms of the Frobenius-Perron operator. Finally, we introduce the 
Ruelle-Pollicott resonances $\lambda_\nu$ which are to identify as decay rates,  
\begin{equation}
C(n)=\lb A\mb {\cal P}^n\mb f\rb=\sum_\nu a_\nu \lambda_\nu^n
\,,
\end{equation}
where the $a_\nu$ denote the coefficients of the resonance expansion. 
We here assume the simplest case that resonances appear with multiplicity $1$, 
otherwise the spectral decomposition of the Frobenius-Perron operator would be given 
by a so-called Jordan block structure, whereby the expansion on the rhs 
would become more complicate. 
It should be remarked that the decay rate is precisely given by the logarithm of 
$\lambda_\nu$ which is more convenient to consider in continuous-time dynamics. 
While the $\lambda_\nu$ are located inside the unit circle on the complex plane, 
the logarithm of $\lambda_\nu$ are customarily chosen to be in the lower half plane. 
Due to the fact that the Frobenius-Perron operator preserves the positivity 
of density functions the resonances are real or appear as complex pairs. 
The trace of the Frobenius-Perron operator defined through 
${\rm tr}{\cal P}^n=\int{\rm d}q\,{\rm d}p\,\delta^2((q,p)-{\cal M}^n(q,p))$, 
i.e. setting image and original points in (\ref{fpoper}) equal, becomes 
\begin{equation}
{\rm tr}{\cal P}^n=1+\sum_\nu\lambda_\nu^n
\label{trres}
\end{equation}
in terms of the resonances. 
We here have separated the stationary eigenvalue $1$ from the summation 
of the resonances. 
We should carefully distinguish between forward and backward time evolution, 
since we don't expect an increase of correlations for the backward time propagation. 
It has been shown that the backward time Frobenius-Perron operator 
has the same resonances. 

Ruelle-Pollicott resonances are defined as the poles 
of the resolvent of ${\cal P}$, 
\begin{equation}
R(z)=\frac{1}{z-{\cal P}}
\,.
\end{equation}
The corresponding ``eigenfunctions'' are not square-integrable functions like those 
of the quantum propagator, but distributions. 
It is known that unstable manifolds of periodic orbits 
function as supports of these singular eigenfunctions. 
For the backward time evolution stable and unstable manifolds exchange their role. 

\section{Approximate Resonances}
Supported by arguments of pertubation theory, 
Weber et al. \cite{weber1,weber2} have shown that 
classical Ruelle-Pollicott 
resonances of the Frobenius-Perron operator 
can be found by investigating the propagator restricted on different 
phase-space resolutions. 
Moreover, one finds approximate eigenfunctions which scar along unstable manifolds. 
This important result will be useful for the comparison of classical and quantum eigenfunctions. 

We restrict our considerations on the Hilbert space of square integrable functions $L^2$. 
For the system considered 
the phase space is the unit sphere, where it becomes convenient to use 
the basis of spherical harmonics. 
The Frobenius-Perron operator becomes an infinite unitary matrix 
whose unimodular spectrum can be separated into a discrete part for integrable 
components and into a continuous part for hyperbolic components of the dynamics. 
Truncation to an $M\times M$ matrix 
corresponding to a restricted phase-space resolution destroys 
unitarity. The spectrum becomes discrete and the eigenvalues are 
inside or on the unit circle. 
As $M$ increases some eigenvalues prove $M$-independent; 
these are said to be stabilized. 
Stabilized eigenvalues reflect spectral properties of 
the Frobenius-Perron operator. They are (almost) unimodular for 
integrable components or stable islands. 
In contrast, eigenvalues which 
are stabilized inside the unit circle reflect Ruelle-Pollicott resonances. 
Non-stabilized eigenvalues have typically smaller moduli 
than the stabilized ones. 
They change their positions as $M$ increases till they reach positions of 
resonances where they can settle for good.
We also know that the eigenfunctions corresponding to stabilized eigenvalues 
are localized either on tori for unimodular eigenvalues or around 
unstable manifolds for non-unimodular eigenvalues. We 
expect that the latter eigenfunctions 
converge weakly to singular resonance ``eigenfunctions''. 
We will call them approximate resonance eigenfunctions in the sequel.

\begin{figure}
\begin{center}
\leavevmode
\epsfxsize=1\textwidth
\epsffile{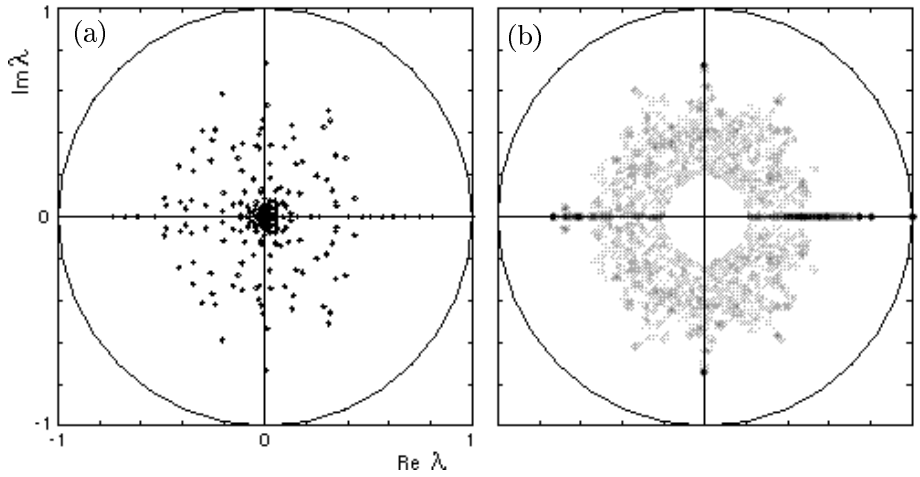}
\end{center}
\caption{
(a) Eigenvalues of the truncated Frobenius-Perron matrix ($l_{\rm max}=32$). 
(b) Eigenvalue density computed from Frobenius-Perron matrices with 
$l_{\rm max}=20,21,\dots 70$. The centered disc is not shown, 
because of the increasing density at the origin (see (a)).}
\label{resdens}
\end{figure}

\begin{figure}
\begin{center}
\leavevmode
\epsfxsize=1\textwidth
\epsffile{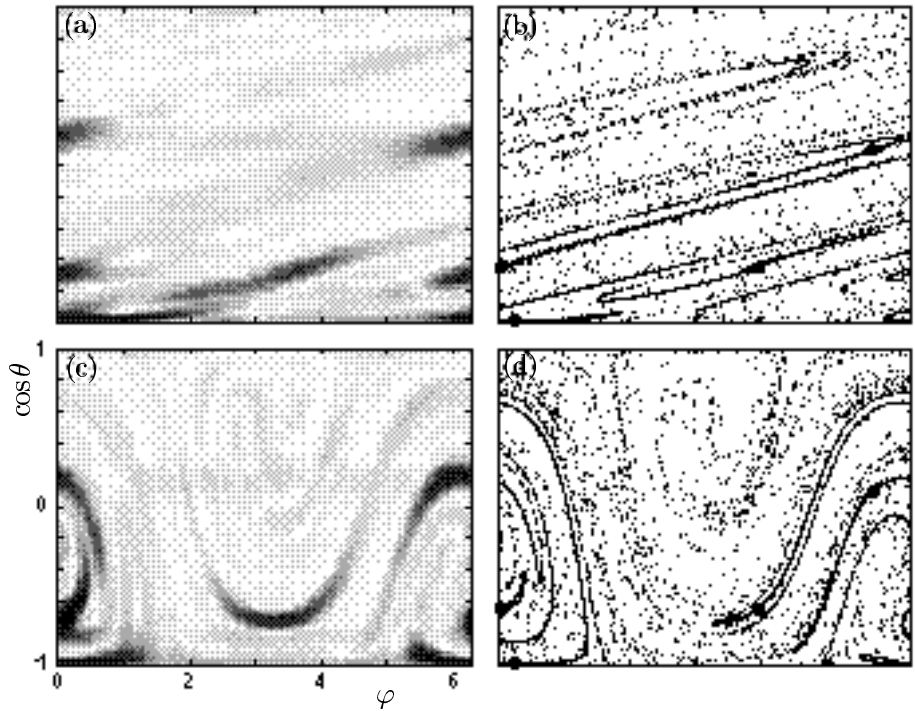}
\end{center}
\caption{(a) Grey-shade coded phase-space plot of the modulus of the approximate 
resonance eigenfunction ($l_{\rm max}=32$) 
from $\lambda\approx -{\rm i}0.75$ (No~4 in Tab.~\ref{restab}). 
(b) Unstable manifold of a weakly unstable period-4 orbit 
(spots) supporting the resonance eigenfunction. 
(c) Approximate resonance eigenfunction of backward time evolution from same resonance as in (a). 
(d) Stable manifold of the same orbit as in (b).}
\label{ef4}
\end{figure}

Fig.~\ref{resdens} (a) shows the eigenvalues of the truncated 
Frobenius-Perron matrix in the complex plane. 
Here the dimension is $M=(l_{\rm max}+1)^2$, where $l_{\rm max}$ is the 
maximal total angular momentum of the spherical harmonics in which the 
density functions are expanded. 
In (b) we see a grey-scale shaded plot of the eigenvalue density calculated 
from truncated matrices ($l_{\rm max}=20, 21, \dots, 70$). 
Dark spots corresponding to large amplitudes of the density indicate 
resonance positions, since resolution independent eigenvalues highly increase 
the density through accumulation.  
A comparison of both shows that some eigenvalues in (a) reflect resonance positions
(see also Tab.~1). 
In Fig.~\ref{ef4} (a) the modulus of the approximate eigenfunction is plotted in phase space, 
where the dark regions belong to large moduli of the complex-valued function. 
The comparison with the unstable manifolds (b) of a weakly unstable period-4 orbit 
shows that the approximate eigenfunction scars along these manifolds. 
In (c) and (d) we see the scarring of the eigenfunction of backward time propagation 
(same resonance as in (a)) along the stable manifolds.

\section{Coherent-State Representation}
To represent quantum operators in a way close the corresponding 
classical observables it is convenient to start from coherent states. 
For the $SU(2)$ group 
coherent states can be generated by a rotation of the 
state $|j,m=j\rangle$ as $|\theta,\varphi\rangle=R(\theta,\varphi)|jj\rangle$. 
In the $|jm\rangle$ basis these coherent states are given by 
\begin{equation}
\label{cs}
|\theta,\varphi\rangle=\left(1+\left(\tan\frac{{}_\theta}{{}^2}\right)^2\right)^{-j}
\sum_{m=-j}^j\sqrt{2j\choose j-m}
\left(\tan\frac{{}_\theta}{{}^2}\,{\rm e}^{{\rm i}\varphi}\right)^{j-m}
|jm\rangle
\,.
\end{equation}
Since the set of the coherent states is overcomplete, there are several 
ways describing quantum operators with coherent states \cite{qfunk,glauber,perelomov}. 
On the one hand, one can use the so-called $P$ function, defined 
as the weight of coherent-state projectors in the continuous mixture 
\begin{equation}
A=\frac{N}{4\pi}
\int{\rm d}\Omega \,
P_A(\theta,\varphi)\,
|\theta,\varphi\rangle\langle\theta,\varphi|
\,.
\end{equation}
The integral is over the unit sphere, 
where ${\rm d}\Omega={\rm d}\varphi\,{\rm d}\theta\,\sin\theta$. 
On the other hand, we have the so-called $Q$ function,  
the coherent state expectation value of an operator, 
$Q_A(\theta,\varphi)=\langle\theta,\varphi|A|\theta,\varphi\rangle$. 
It is important that $Q$ is a smooth function on 
phase space, 
while $P$ can strongly oscillate, 
particularly in the shortest wave lengths. 
However, in contrast to the coherent states of Weyl groups, $P$ functions 
always exist. Both functions can be expanded in terms of
spherical harmonics \cite{online}, 
\begin{eqnarray}
Q_A(\theta,\varphi)&=&\sum\limits_{l=0}^{2j}\sum\limits_{m=-l}^l
q_{lm}(A)Y_l^m(\theta,\varphi) \nonumber\\
P_A(\theta,\varphi)&=&\sum\limits_{l=0}^{2j}\sum\limits_{m=-l}^l
p_{lm}(A)Y_l^m(\theta,\varphi)
\,,
\end{eqnarray}
where the summations break off at the total angular momentum $l=2j$. 
The Hilbert space of phase-space functions 
must not be confused with the Hilbert space of 
quantum wave functions; 
we therefore use round brackets 
for the scalar product already introduced in (\ref{average}), 
$\lb\, f\mb\, g\rb=
\int{\rm d}\Omega\,f^\ast(\theta,\varphi)g(\theta,\varphi)  $. 
Although the $P$ functions tend to oscillate more strongly than 
the $Q$ functions, the expansion coefficients 
of the $P$ and $Q$ functions corresponding to the same operator converge 
to one another in the classical limit, $p_{lm}(A)\jinft q_{lm}(A)$ 
for fixed $l,m$. 
It is easy to see that the trace of an operator product can be written as 
the scalar product of $P$ and $Q$ as 
${\rm tr}A^\dagger B=\frac{N}{4\pi}\lb P_A\mb Q_B\rb$. 
In particular, if the set $\{|\phi_k\rangle\}$ of wave functions 
form an orthogonal basis of the quantum Hilbert space, 
then the $P$ and $Q$ functions of ket-bras 
$P_{ik}\equiv P_{|\phi_i\rangle\!\langle\phi_k|}$ 
and $Q_{ik}\equiv Q_{|\phi_i\rangle\!\langle\phi_k|}$ generate 
biorthonormal sets in the Hilbert space of phase-space functions
\begin{equation}
\frac{N}{4\pi}\lb P_{ik}\mb Q_{i'k'}\rb={\rm tr}
\left(|\phi_k\rangle\!\langle\phi_i|
|\phi_{i'}\rangle\!\langle\phi_{k'}|\right)
=\delta_{kk'}\delta_{ii'}
\,.
\end{equation}
The $Q$ function of a density operator $Q_\rho$ is also called 
Husimi function. If the density operator is a projector of 
form $|\psi\rangle\langle\psi|$ the corresponding Husimi function 
is a phase-space representation of a quantum wave function. 
We call the $Q_{ik}\equiv Q_{|\phi_i\rangle\!\langle\phi_k|}$ 
Husimi eigenfunction if the $|\phi_i\rangle$ denote Floquet eigenstates. 
The latter notation becomes obvious in the next section. 
We further distinguish between diagonal Husimi eigenfunctions $Q_{kk}$ 
and skew ones $Q_{ik}$ with $i\ne k$.

\section{Husimi Propagator}
The Husimi propagator ${\cal F}$ is defined through the time evolution of 
a quantum density operator $\rho$ as
\begin{equation}
Q_{\rho(n)}(\theta,\varphi)=
\langle\theta,\varphi|F^n\rho (0)(F^{\dagger})^n|\theta,\varphi\rangle
={\cal F}^nQ_{\rho(0)}(\theta,\varphi)
\,.
\end{equation}
Using the Floquet eigenstates, 
$F|\phi_i\rangle={\rm e}^{-{\rm i}\phi_i}|\phi_i\rangle$, 
the Husimi eigenfunctions are easily calculated as 
$Q_{ik}\equiv\langle\theta,\varphi|\phi_i\rangle\langle\phi_k|\theta,\varphi\rangle$. 
The Husimi propagator thus has $N^2$ unimodular eigenvalues 
whose phases are differences of the Floquet eigenphases. 
There is an $N$-fold degeneracy of the eigenvalue $1$ corresponding to 
the diagonal Husimi eigenfunctions $Q_{kk}$ which 
are real and normalized as $\int{\rm d}\Omega\,Q_{kk}=4\pi/N$. 
All other (skew) Husimi eigenfunctions with $i\ne k$ are complex 
and their phase space-integral vanishes. 
In basis of spherical harmonics the Husimi propagator 
becomes an $N^2\times N^2$ matrix (Husimi matrix). 
The diagonal representation of the 
Husimi propagator is simply given as
\begin{equation}
{\cal F}^n=\frac{2j+1}{4\pi}\sum_{i=0}^{2j}
\sum_{k=0}^{2j}\mb Q_{ik}\rb 
{\rm e}^{-{\rm i}n(\phi_i-\phi_k)}\lb P_{ik} \mb
\,.
\label{diagrep}
\end{equation}
The Husimi spectral density is identified as the density-density 
correlation function with respect to the Floquet eigenphases, 
\begin{equation}
C(\omega)=\int_0^{2\pi}{\rm d}\phi \rho(\omega+\phi)\rho(\phi)
=\sum_{ik}\delta(\omega-(\phi_i-\phi_k))
\,,
\end{equation}
where $\rho(\phi)=\sum_k\delta(\phi-\phi_k)$. 
Some authors prefer the normalized density 
$\tilde{\rho}(\omega)=N^{-1}\sum_k\delta(\phi-\phi_k)$.

\section{$L^2$ Norms of Husimi Eigenfunctions}
The squared $L^2$ norm of a diagonal Husimi eigenfunction, 
\begin{equation}
\label{l2normdiag}
||Q_{kk}||^2=\int{\rm d}\Omega\,|\langle 
\theta,\varphi|\phi_k\rangle|^4
\,,
\end{equation} 
is the inverse participation ratio (IPR) 
with respect to coherent states (phase-space distribution). 
It becomes large if the Husimi function is strongly 
localized in phase space, say scarred on periodic orbits. 
On the other hand, the squared $L^2$ norm of a skew Husimi eigenfunction 
can be understood as the overlap of two diagonal Husimi functions 
on phase space, 
\begin{equation}
\label{l2normskew}
\int{\rm d}\Omega\,|Q_{ik}|^2=\int{\rm d}\Omega\,|\langle 
\theta,\varphi|\phi_i\rangle|^2|\langle\theta,\varphi|\phi_k\rangle|^2
\,.
\end{equation}
From the Schwarz inequality, 
\begin{equation}
||Q_{ik}||^2\le ||Q_{ii}||\,||Q_{kk}||
\,,
\end{equation}
it becomes obvious that for large values of $||Q_{ik}||^2$ 
both diagonal Husimi eigenfunctions 
must be localized in the same phase-space regions. 

We illustrate two examples. 
A constant function on the sphere is of course a uniformly distributed function. 
But note that it is not a Husimi eigenfunction, since the corresponding density 
operator is of form $\rho=\frac{1}{N}{\bf 1}$. 
Using the normalization 
$\frac{N}{4\pi}\int{\rm d}\Omega\,Q_{\rho}=1 $ 
the Husimi function becomes $\frac{1}{N}$. For the squared $L^2$ norm we find 
\begin{equation}
||Q_{\rho}||^2=\frac{4\pi}{N^2}
\end{equation}
for uniform function. 
In comparison, the random-matrix averaged squared $L^2$ norms of 
diagonal Husimi eigenfunctions is larger by a factor 2 (see Sec.~\ref{RMT})
which can be explained by quantum fluctuations. 
A most strongly localized Husimi function, in contrast, corresponds to a density oparator 
which is a coherent-state projector. 
Due to the invariance under rotations the $L^2$ norm is the same for all 
coherent-state Husimi functions. Using the coherent-state projector 
$|jj\rangle\langle jj|$ and 
(\ref{exaktl2}) one easily finds for the squared $L^2$ norm 
\begin{equation}
||Q_{|jj\rangle\!\langle jj|}||^2=\frac{2\pi}{N}
\end{equation}
for strongly localized Husimi function. 
It becomes obvious that squared $L^2$ norms of 
most strongly localized and uniformly distributed 
Husimi functions differ by a factor of order $N$.

We here show the 
calculation of the $L^2$ norm from vector coefficients 
in the $|jm\rangle$ basis ($c^k_m=\langle jm|\phi_k\rangle$). 
Introducing the decomposition of unity in terms of the angular 
momentum states, ${\bf 1}=\sum_{m=-j}^j |jm\rangle\langle jm|$, 
the $L^2$ norm of a skew Husimi eigenfunction 
becomes a four-fold sum, 
\begin{eqnarray}
\int{\rm d}\Omega\,
|\langle\theta,\varphi |\phi_i\rangle\langle\phi_k|\theta,\varphi\rangle|^2
&=&\sum_{m_1\dots m_4}
c^i_{m_1}(c^i_{m_2})^\ast c^k_{m_3}(c^k_{m_4})^\ast
\nonumber\\ &&\hspace{-2cm}\times
\int{\rm d}\Omega\,
\langle\theta,\varphi |jm_1\rangle \langle jm_2|\theta,\varphi\rangle
\langle\theta,\varphi |jm_3\rangle \langle jm_4|\theta,\varphi\rangle
\label{summation}
\,.
\end{eqnarray}
From the definition (\ref{cs}) of the coherent states we find, 
using the relations $\tan\frac{\theta}{2}=\frac{1-\cos\theta}{\sin\theta}
=\frac{\sin\theta}{1+\cos\theta}$ for $0\le \theta\le\pi$, 
\begin{equation}
\langle jm|\theta,\varphi\rangle=2^{-j}\sqrt{2j\choose j-m}
(1-\cos\theta)^\frac{j-m}{2}(1+\cos\theta)^\frac{j+m}{2}{\rm e}^{{\rm i}(j-m)\varphi}
\,.
\end{equation}
After a few further steps the phase-space integral 
$\int{\rm d}\Omega\,=\int_0^{\pi}{\rm d}\theta\,\sin\theta\int_0^{2\pi}{\rm d}\varphi$ 
leads to a Kronecker $\delta$ and to a beta function \cite{gradshteyn} for the $\varphi$- and 
$\theta$-dependent part, respectively. We thus obtain for the coefficients of the 
summation (\ref{summation})
\begin{eqnarray}
&&\int{\rm d}\Omega\,
\langle\theta,\varphi |jm_1\rangle \langle jm_2|\theta,\varphi\rangle
\langle\theta,\varphi |jm_3\rangle \langle jm_4|\theta,\varphi\rangle
\nonumber\\
&&=4\pi\frac{\Gamma(\frac{4j-\sum m_i}{2}+1)\Gamma(\frac{4j+\sum m_i}{2}+1)}
{\Gamma(4j+2)}
\nonumber\\ &&\times
\delta(m_1\!-\!m_2\!+\!m_3\!-\!m_4)
\sqrt{{2j\choose j-m_1}{2j\choose j-m_2}{2j\choose j-m_3}{2j\choose j-m_4}}
\label{sumcoeff}
\end{eqnarray}
We make use of the argument of the 
Kronecker $\delta$ and get for the $L^2$ norm
\begin{eqnarray}
\int{\rm d}\Omega\,
|\langle\theta,\varphi |\phi_i\rangle\langle\phi_k|\theta,\varphi\rangle|^2
&=&4\pi\sum_{m_1\dots m_4}\frac{(2j-m_1-m_3)!(2j+m_1+m_3)!}{(4j+1)!}
\nonumber\\
&\times&
\sqrt{{2j \choose j-m_1}{2j \choose j-m_2}
{2j \choose j-m_3}{2j \choose j-m_4}}
\nonumber\\ &\times&
\delta(m_1\!-\!m_2\!+\!m_3\!-\!m_4)
c^i_{m_1}(c^i_{m_2})^\ast c^k_{m_3}(c^k_{m_4})^\ast
\,.
\nonumber\\
\label{exaktl2}
\end{eqnarray}
The latter formular will be used for the calculation of Husimi $L^2$ norms 
of random vectors in the next section. 

\section{Random-Matrix Average of $L^2$ Norms}
\label{RMT}
In order to compare the results of the kicked top with those of random-matrix theory 
we now consider diagonal Husimi eigenfunctions and replace 
the coefficients by real or complex random numbers due to 
the orthogonal (COE) or unitary ensemble (CUE), respectively \cite{book}. 
It is easy to see from (\ref{l2normdiag},\ref{l2normskew}) that the $L^2$ norm of a Husimi 
function is invariant under rotations. As it has been shown in Sec.~\ref{kickedtop} 
the eigenvectors of the kicked top must be compared to real random vectors.  
The only correlation between the coefficients is the 
normalization of the vector, $\sum_m |c_m|^2=1$ 
(we drop the upper index in the following). 
Therefore we neglegt all terms containing random phases and 
keep the contributing terms with $m_1=m_2\wedge m_3=m_4$ or 
$m_1=m_4\wedge m_2=m_3$. 
By the choice of real coefficients one has the further possibility 
$m_1=m_3\wedge m_2=m_4$. 
We may abbreviate the coefficients of the summation (\ref{sumcoeff}) 
as $f(m_1,\dots,m_4)$ for a moment to find out the contributing terms,
\begin{eqnarray}
\left\langle||Q_{kk}||^2\right\rangle&=&
\sum_{m_1\dots m_4}
f(m_1,\dots,m_4)\,\delta(m_1\!-\!m_2\!+\!m_3\!-\!m_4)
\left\langle c_{m_1}c_{m_2}^\ast c_{m_3}c_{m_4}^\ast\right\rangle
\nonumber\\
&=&\sum_{m_1\dots m_4}
f(m_1,\dots,m_4)\,\delta(m_1\!-\!m_2\!+\!m_3\!-\!m_4)
\nonumber\\&\times&\!\!
\bigg(\Big[\delta_{m_1m_2}\delta_{m_3m_4}(1-\delta_{m_1m_3})
+\delta_{m_1m_4}\delta_{m_2m_3}(1-\delta_{m_1m_3})
\nonumber\\
&+&\!\!\{\delta_{m_1m_3}\delta_{m_2m_4}(1-\delta_{m_1m_2})\}\Big]
\left\langle|c_{m}|^2|c_n|^2\right\rangle
+\delta_{m_im_k}^3\left\langle|c_m|^4\right\rangle\!\bigg)
\,.
\nonumber\\
\end{eqnarray}
The diagonal part $\delta_{m_im_k}^3= \delta_{m_1m_2}\delta_{m_2m_3} \delta_{m_3m_4}$ 
is separated, since there 
appears the average of $|c_m|^4$. Due to the symmetry of $f(m_1,\dots,m_4)$ 
the first two terms in the bracket give the same contribution; to that end, 
one executes the summations over $m_2$ and $m_4$. 
The contribution coming from real coefficients $\{\dots\}$ vanishes; 
here one summates over $m_3$ and $m_4$. Finally, one finds for 
the averaged $L^2$ norm
\begin{eqnarray}
\left\langle||Q_{kk}||^2\right\rangle&=&
\frac{4\pi}{4j+1}\bigg[2\sum_{m,n=-j\atop m\ne n}^j{4j\choose 2j-m-n}^{-1}
{2j\choose j-m}{2j \choose j-n}
\left\langle|c_m|^2|c_n|^2\right\rangle
\nonumber\\
&+&\sum_{m=-j}^j{4j \choose 2j-2m}^{-1}{2j\choose j-m}^2
\left\langle|c_m|^4\right\rangle\bigg]
\,.
\label{sumbinom}
\end{eqnarray}
Next we need the averages of products,
\begin{eqnarray}
\left\langle |c_m|^2|c_n|^2\right\rangle&=&
\left\{ \begin{array}{cc}
\frac{1}{N(N+1)} & ,\,{\rm CUE}\\
\frac{1}{N(N+2)} & ,\,{\rm COE}
\end{array}\right.
\label{avprd}
\,,\\
\left\langle |c_m|^4\right\rangle&=&
\left\{ \begin{array}{cc}
\frac{2}{N(N+1)} & ,\,{\rm CUE}\\
\frac{3}{N(N+2)} & ,\,{\rm COE}
\end{array}\right.
\,.
\label{avipr}
\end{eqnarray}
The averages are easily calculated from the probability distribution given in \cite{book}. 
It should be remarked here that semiclassical corrections are of next to leading order of $N^{-1}$. 
Therefore we do not neglect this order in our RMT results.  

Let us consider first the CUE. 
Since $\left\langle|c_m|^4\right\rangle=2\left\langle |c_m|^2|c_n|^2\right\rangle$, 
we can complete the summations over the diagonal $(m=n)$ and off-diagonal $(m\ne n)$ part. 
Thus we have for the CUE
\begin{equation}
\left\langle||Q_{kk}||^2\right\rangle=
\frac{8\pi}{N(N+1)}\frac{1}{4j+1}\sum_{m,n=-j}^j{4j\choose 2j-m-n}^{-1}
{2j\choose j-m}{2j \choose j-n}
\,.
\end{equation}
It might not be easy to see that this results to $\left\langle||Q_{kk}||^2\right\rangle=\frac{8\pi}{N(N+1)}$. 
To that end we present an alternative way to calculate the averaged $L^2$ norm which 
is unfortunately not applicable for the COE \cite{gnutz}. 
An arbitrary complex random vector can be written as $|\psi\rangle=U|jj\rangle$, 
where $U$ is a unitary random matrix. 
A coherent state is generated through a rotation operator as $|\theta,\varphi\rangle=R(\theta,\varphi)|jj\rangle$. 
We thus have $\langle\theta,\varphi|\psi\rangle=\langle jj|R^\dagger U|jj\rangle$. 
The product $R^\dagger U=\tilde{U}$ defines a new unitary random matrix, where the Haar measure 
keeps unchanged ${\rm d}\mu(\tilde{U})={\rm d}\mu(U)$. 
Therefore the averaged $L^2$ norm becomes 
\begin{eqnarray}
\left\langle||Q_{kk}||^2\right\rangle&=&\int{\rm d}\mu(U)\int{\rm d}\Omega\,|\langle jj|R^\dagger U|jj\rangle|^4
\nonumber\\
&=&\int{\rm d}\Omega\int{\rm d}\mu(\tilde{U})\,|\langle jj|\tilde{U}|jj\rangle|^4
\nonumber\\
&=&4\pi\left\langle|c_m|^4\right\rangle
\,.
\end{eqnarray}
Note that for the COE, in contrast, one deals with real vectors which 
generally become complex after rotating.

For the COE we again complete the diagonal and off-diagonal summations 
in (\ref{sumbinom}), whereby 
a further contribution remains, because of the factor $3$ of the  fourth moment (\ref{avipr}), 
\begin{equation}
\frac{4\pi}{4j+1}\sum_{m=-j}^j{4j \choose 2j-2m}^{-1}{2j\choose j-m}^2
\frac{1}{N(N+2)}
=\frac{(4^{j}(2j)!)^2}{(4j+1)(4j)!}\,\frac{4\pi}{N(N+2)}\;\;\;
\label{forcontrib}
\end{equation}
The latter equation can be calculated as follows \cite{pb}. We rewrite 
\begin{equation}
\sum_{m=-j}^j{4j \choose 2j-2m}^{-1}{2j\choose j-m}^2
=\frac{(2j)!^2}{(4j)!}
\sum_{k=0}^{2j}{2k\choose k}{4j-2k\choose 2j-k}
\end{equation}
with $k=m+j$. We now show that
\begin{equation}
\sum_{k=0}^n{2k\choose k}{2n-2k\choose n-k}=4^n
\,.
\end{equation}
The generating function of ${ 2k\choose k}$ is 
\begin{equation}
\frac{1}{\sqrt{1-4x}}=\sum_{k=0}^\infty {2k\choose k}x^k
\end{equation}
which is easily seen from Taylor expansion,
\begin{equation}
\frac{1}{k!}\frac{{\rm d}^k}{{\rm d}x^k}(1-4x)^{-\frac{1}{2}}\bigg|_{x=0}=
\frac{1}{k!}\frac{1}{2}\left(1+\frac{1}{2}\right)\dots\left(k-\frac{1}{2}\right)4^k
=\frac{(2k)!}{k!^2}
\,.
\end{equation}
Squaring the generating function gives 
\begin{equation}
\frac{1}{1-4x}=\sum_{k=0}^\infty\sum_{l=0}^\infty 
{2k\choose k}{2l\choose l}x^{k+l}=
\sum_{n=0}^\infty x^n\sum_{k=0}^n{2k\choose k}{2n-2k\choose n-k}
\,,
\end{equation}
where $n=k+l$. 
Comparing the latter equation with the geometric series
\begin{equation}
\frac{1}{1-4x}=\sum_{n=0}^\infty x^n 4^n
\end{equation}
finishes the proof. 

The aforementioned contribution (\ref{forcontrib}) becomes 
more familiar by the approximation
\begin{equation}
\frac{(4^{j}(2j)!)^2}{(4j+1)(4j)!}=
\frac{\sqrt{\pi}}{2}\frac{\Gamma(2j+1)}{\Gamma(2j+3/2)}
\approx \frac{1}{2}\sqrt{\frac{\pi}{2j+1}}
\,.
\end{equation}
The averaged $L^2$ norms finally become
\begin{equation}
\left\langle ||Q_{kk}||^2\right\rangle\simeq\frac{4\pi}{N^2} 
\left\{ \begin{array}{cc}
\frac{2N}{N+1} & ,\,{\rm CUE}\\
\frac{N}{N+2}\left(2+\frac{1}{2}\sqrt{\frac{\pi}{N}}\right) & ,\,{\rm COE}
\end{array}\right.
\,,
\end{equation}
As an interesting point to remark here the COE eigenstates are somewhat 
more localized than the CUE eigenstates, since their averaged $L^2$ norm 
contains a further contribution 
of order $N^{-\frac{5}{2}}$. 
But in contrast to the fourth moment (\ref{avipr}) - it might be understood as the 
averaged IPR with respect to the $|jm\rangle$ basis - the difference vanishes in 
the classical limit $N\rightarrow \infty$. 
However, for both ensembles the averaged squared $L^2$ norms are roughly 
twice larger than for a constant distribution on the sphere. 

For the skew Husimi eigenfunctions the averaged $L^2$ norms can be 
calculated from the relation
\begin{equation}
4\pi=\int{\rm d}\Omega\,|\langle\theta,\varphi|\theta,\varphi\rangle|^2=
\int{\rm d}\Omega\,\left(\sum_k|\langle\theta,\varphi|\phi_k\rangle|^2\right)^2
=\sum_k ||Q_{kk}||^2+\sum_{i\ne k}||Q_{ik}||^2
\,,
\end{equation}
which leads to
\begin{equation}
\left\langle||Q_{ik}||^2\right\rangle=
\frac{4\pi-N\left\langle||Q_{kk}||^2\right\rangle}{N(N-1)}
\,.
\end{equation}
Introducing the averages of the diagonal ones we find for CUE 
\begin{equation}
\left\langle||Q_{ik}||^2\right\rangle=\frac{4\pi}{N^2}\frac{N}{N+1}
\,.
\end{equation}
For the COE one obtaines the same result up to an order of $N^{-\frac{7}{2}}$, 
\begin{equation}
\left\langle||Q_{ik}||^2\right\rangle\simeq\frac{4\pi}{N^2}\frac{N^2}{(N-1)(N+2)}
\left(1-\frac{1}{2N}\sqrt{\frac{\pi}{N}}\right)
=\frac{4\pi}{N^2}\frac{N}{N+1}+{\cal O}(N^{-\frac{7}{2}})
\label{skewl2coe}
\,.
\end{equation}

\section{Comparison of Quantum and Classical Eigenfunctions}
\begin{figure}
\begin{center}
\leavevmode
\epsfxsize=1\textwidth
\epsffile{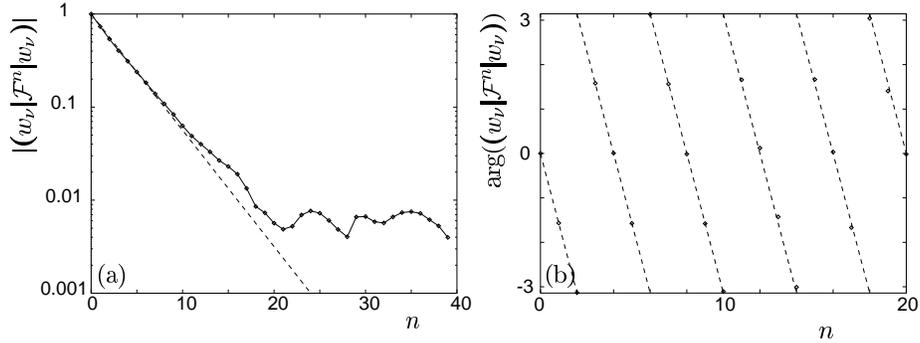}
\end{center}
\caption{Return probability of the approximate resonance eigenfunction from 
$\lambda\approx-{\rm i}0.75$ (No~4 in Tab.~\ref{restab}; see also Fig.~\ref{ef4} (a)): 
(a) the amplitude shows for small $n$ an exponential decay of $(0.75)^n$ (dashed), while 
the phase (b) evolves like $-n\frac{\pi}{2}$.}
\label{lambda1r}
\end{figure}

It has been shown in \cite{weber3} that 
quantum quasiprobability propagation looks classical if 
phase-space resolution is blurred such that 
Planck cells are far from being resolved. 
As a result, if we truncate the Husimi matrix to coarse 
phase-space resolution, then it becomes almost equal to the truncated 
Frobenius-Perron operator, $T{\cal F}T\jinft T{\cal P}T$, 
where 
\begin{equation}
T=\sum_{l=0}^{l_{\rm max}}\sum_{m=-l}^l\mb Y_l^m\rb\lb Y_l^m\mb 
\end{equation}
denotes the truncation projector restricting the ``classical'' Hilbert space dimension 
to $M=(l_{\rm max}+1)^2$. 
In the following we choose $M\ll N^2$, 
and thus have $T{\cal F}T\approx T{\cal P}T$. 
Again we consider the classical propagator. 
Let $\mb w\rb=T\mb w\rb$ be an approximate resonance eigenfunction 
with stabilized eigenvalue $\lambda$ of $T{\cal P}T$. 
Since the stabilized eigenvalues reflect spectral properties 
of the non-truncated Frobenius-Perron operator, $\lambda^n$ 
must be a stabilized eigenvalue of $T{\cal P}^nT$, at least 
if $M$ is chosen large enough. This property does not hold 
for non-stabilized eigenvalues, since 
$T{\cal P}^nT\ne (T{\cal P}T)^n$. 
Choosing the approximate resonance eigenfunction $L^2$ normalized, 
the return probability becomes $\lb w\mb {\cal P}\mb w\rb=\lambda^n$ 
for small $n$ (see Fig.~\ref{lambda1r}). 
Note that the latter property makes sense only if we consider approximate 
resonance eigenfunctions, because the singular resonance eigenfunctions are 
not square integrable. 
Fig. \ref{lambda1r} shows the return probability of the approximate eigenfunction 
from stabilized eigenvalue $\lambda_\nu\approx -{\rm i}0.75$ (No 4 in Tab. \ref{restab}). 
As well the moduli as the phases coincide with $\lambda_\nu^n$ for about 20 iterations. 
Beyond this time quantum fluctuations become visible. 

\begin{figure}
\begin{center}
\leavevmode
\epsfxsize=1\textwidth
\epsffile{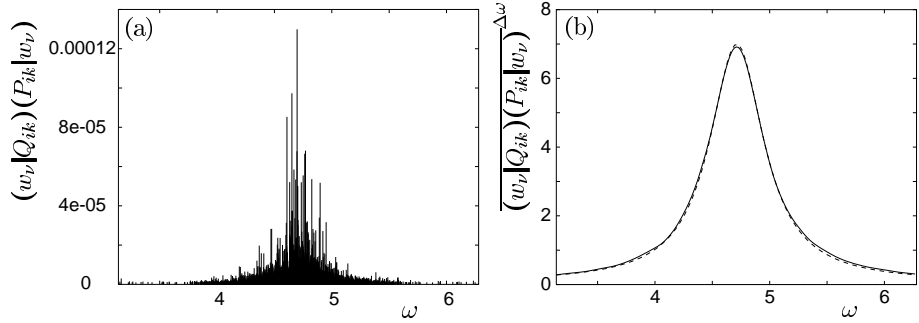}
\end{center}
\caption{(a) Overlaps of Husimi eigenfunctions with the approximate resonance eigenfunction 
from $\lambda\approx -{\rm i}0.75$ (No~4 in Tab.~\ref{restab}). 
(b) Smoothed overlaps (solid) in comparison with the Lorentzian distribution (dashed) 
corresponding to $\lambda_4=-{\rm i}0.75$ 
(see text).}
\label{lambda1f}
\end{figure}

Since $T{\cal F}T\approx T{\cal P}T$, we expect 
$T{\cal F}^nT\approx T{\cal P}^nT$ for small $n$. 
Now we can replace the Husimi propagator by its 
diagonal representation (\ref{diagrep}),
\begin{equation}
\lambda^n\approx
\lb w\mb {\cal F}^n\mb w\rb=\frac{N}{4\pi}\sum_{ik}
\lb w\mb Q_{ik}\rb {\rm e}^{-{\rm i}n(\phi_i-\phi_k)}
\lb P_{ik}\mb w\rb
\,.
\end{equation}
The return probability becomes a double sum of overlaps of quantum 
and classical eigenfunctions. For large $N$ the Husimi eigenphases 
are quite dense in the interval $[0,2\pi)$. 
Outside an interval around the Husimi eigenphase $\omega=0$ wherein 
level repulsion of Floquet eigenphases become perceptible 
the spectral density of differences $\phi_i-\phi_k$ is almost constant, $N^2/2\pi$. 
It is 
convenient to replace the sum by a continuous integral over the 
Husimi eigenphases, where the overlaps can be replaced by a continuous 
function as a smoothed distribution of overlaps,
\begin{equation}
\lambda^n\approx \frac{N}{4\pi}\int\limits_0^{2\pi}{\rm d}\omega\,
\overline{\lb w\mb Q_{ik}\rb\lb P_{ik}\mb w\rb}^{\Delta\omega}(\omega)
{\rm e}^{-{\rm i}n\omega}\,.
\end{equation}
%
In the classical limit we let first $N\longrightarrow\infty$ and 
then $M\longrightarrow\infty$. In this limit the foregoing result becomes 
valid for all $n$. The smoothed overlaps are given by 
the inverse Fourier transform of $\lambda^n$ as 
\begin{equation}
\overline{\lb w\mb Q_{ik}\rb\lb P_{ik}\mb w\rb}
^{\Delta\omega}(\omega)\jinft
\frac{2}{N}\frac{1-|\lambda|^2}{1+|\lambda|^2-
2|\lambda|\cos(\omega-\arg\lambda)}
\label{lorentz}
\,,
\end{equation}
which is a periodic ``Lorentzian'' displaced by the phase 
of $\lambda$ and of width $-\ln|\lambda|$ \cite{keating} (see Fig.~\ref{lambda1f}). 

The smoothing is done by a convolution with a sinc function which naturally 
results from a truncated Fourier transform. 
Here the time restriction is ${\cal N}\le n\le {\cal N}$, where ${\cal N}\approx 20$ 
corresponding to the validity of the return probability 
of the approximate eigenfunction (Fig.~\ref{lambda1f}). 
For simplification the integral is restricted on the interval between the 
first zeros of the sinc function, 
\begin{equation}
\overline{f}^{\Delta\omega}(\omega)\propto
\int\limits_{\omega-\frac{\pi}{\cal N}}^{\omega+\frac{\pi}{\cal N}}
{\rm d}\omega'\frac{\sin ({\cal N}\omega')}{\omega'}f(\omega-\omega')
\,.
\label{smooth}
\end{equation}

\begin{figure}
\begin{center}
\leavevmode
\epsfxsize=1\textwidth
\epsffile{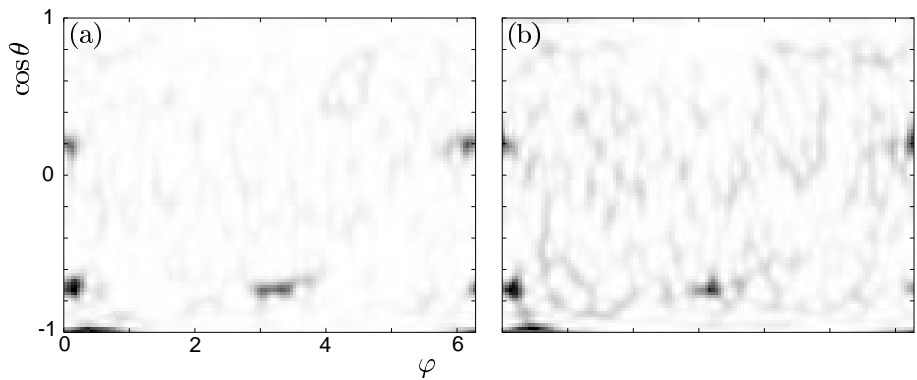}
\end{center}
\caption{Grey-shade coded phase-space plots of the diagonal Husimi eigenfunctions 
(a) No~30 and (b) No~11 of Tab.~\ref{hustab}. 
The Husimi function in (a) is strongly scarred 
in phase-space regions, where the approximate resonance eigenfunctions 
are scarred (Fig.~\ref{ef4}). 
The Husimi function in (b) is also scarred in the same phase-space regions, but 
shows more structures all over the phase space than the function in (a).}
\label{husp4}
\end{figure}

A simple argument connects quantum scars with Ruelle-Pollicott resonances. 
Classical resonance eigenfunctions are scarred along unstable 
manifolds (or stable manifolds for backward time propagation). 
Quantum eigenfunctions which strongly overlap with 
resonance eigenfunctions have to be scarred as well. 
This is an important result, because it explains scarring 
of quantum eigenfunctions not only on periodic orbits, but also 
along stable and unstable manifolds. 
This has been observed for the kicked top \cite{saracenotop}. 
For instance, we consider the skew Husimi eigenfunction which shows the 
largest overlap with the classical approximate eigenfunction considered before. 
The corresponding diagonal Husimi eigenfunctions (No~11 and No~30 in Tab.~\ref{hustab}) 
plotted in Fig.~\ref{husp4} are scarred in the same phase-space regions, 
where the the approximate resonance eigenfunctions are scarred (Fig.~\ref{ef4}), while the 
difference of their Floquet eigenphases ($\phi_{30}-\phi_{11}=-1.586$) is close to the 
phase of the resonance ($\arg\lambda_4=-\frac{\pi}{2}$).

\section{Resonance Corrections of Averaged Phase-Space Overlaps}
In the foregoing section a qualitative explenation of the connection 
between resonances and quantum eigenfunctions was given, but now 
we are interested in more quantitative results. 
To that end, we consider transition rates of coherent states in the classical 
limit
\begin{equation}
\frac{N}{4\pi}|\langle\theta,\varphi|F^n|\theta',\varphi'\rangle|^2\jinft 
\delta\left((q,p)-{\cal M}^n(q',p')\right)
\label{semicl}
\,.
\end{equation}
The latter relation becomes obvious if one suggests that coherent states 
are wave functions most strongly localized on phase-space points. 
We now consider the return probability and integrate over the phase space, 
\begin{equation}
\frac{N}{4\pi}\int{\rm d}\Omega\,
|\langle\theta,\varphi|F^n|\theta,\varphi\rangle|^2
\jinft \int{\rm d}\Omega\,\delta\left((q,p)-{\cal M}^n(q,p)\right)
\,,
\label{trace}
\end{equation}
where the rhs is the trace of the Frobenius-Perron operator \cite{cvit2}. 
We remark here that the integral on the rhs leads to a sum of contributions 
from periodic orbits, which is an important connection between scars on periodic orbits 
and the results of our paper. 
On the one hand, periodic orbits, in particular the weakly unstable ones, 
contribute to the trace of the Frobenius-Perron operator, 
i.e. influence the resonances. 
On the other hand, scars typically appear around weakly unstable periodic orbits. 
On the lhs of (\ref{trace}) we introduce the diagonal representation of the Floquet operator 
and get
\begin{eqnarray}
\frac{N}{4\pi}\int{\rm d}\Omega\,
|\langle\theta,\varphi|F^n|\theta,\varphi\rangle|^2&=&
\frac{N}{4\pi}\int{\rm d}\Omega\,
\left|\sum_k|\langle\theta,\varphi|\phi_k\rangle|^2{\rm e}^{-{\rm i}n\phi_k}\right|^2
\nonumber\\ &=&
\frac{N}{4\pi}\sum\limits_{ik}||Q_{ik}||^2\,{\rm e}^{-{\rm i}n(\phi_i-\phi_k)}
\,.
\end{eqnarray}
Fourier transformation of the latter expression leads to a sum of 
$\delta$ functions weighted by $L^2$ norms,
\begin{equation}
\sum\limits_{n=-\infty}^{\infty}\frac{{\rm e}^{{\rm i}n\omega}}{2\pi}
\sum\limits_{ik}||Q_{ik}||^2\,{\rm e}^{-{\rm i}n(\phi_i-\phi_k)}
=\sum\limits_{ik}||Q_{ik}||^2\,\delta(\omega-(\phi_i-\phi_k))
\label{kamm}
\,.
\end{equation}
Due to the arguments of the foregoing section we expect that 
for finite $N$ the relation (\ref{semicl}) is valid for finite 
times $|n|\le{\cal N}$. 
Validity of semiclassical methods is guarantied for times up to the 
Ehrenfest time, where the number of fixpoints coincides with 
the number of Planck cells. Thus we identify ${\cal N}$ as 
Ehrenfest time. 
The truncated Fourier transform leads to a sum of 
smoothed $\delta$ functions (\ref{smooth}). 
Using (\ref{trace}) we get
\begin{equation}
\overline{||Q_{ik}||^2}^{\Delta\omega}(\omega)
=\frac{4\pi}{N}
\sum_{n=-{\cal N}}^{\cal N} {\rm tr}{\cal P}^n\frac{{\rm e}^{{\rm i}n\omega}}{2\pi}
\,
\end{equation}
which we may call smoothed $L^2$ norms. 
The next step is to drop the stationary eigenvalue $1$ in the traces of the Frobenius-Perron operator. 
The Fourier transform of this eigenvalue leads to a $\delta$ function in the limit ${\cal N}\rightarrow\infty$ 
which is not a point of interest here. 
In the Husimi representation we identify the eigenvalue $1$ as the sum of squared $L^1$ norms 
of Husimi eigenfunctions. 
This is easily seen from the Husimi matrix in basis of spherical harmonics. 
In the first row and column is only one non-vanishing matrix element $\lb Y_0^0\mb {\cal F}\mb Y_0^0\rb=1$. 
Note that $Y_0^0$ is a constant function on the sphere and therefore proportional to the stationary density, 
i.e. it is the eigenfunction from eigenvalue $1$. 
Introducing the diagonal representation of the Husimi propagator (\ref{diagrep}) 
one identifies $\frac{N}{4\pi}\sum_k\lb Y_0^0\mb Q_{kk}\rb\lb P_{kk}\mb Y_0^0\rb =1$. 
Note that $\lb Y_0^0\mb Q_{ik}\rb =\lb P_{ik}\mb Y_0^0\rb =\frac{\sqrt{4\pi}}{N}\delta_{ik}$, where 
$\lb Y_0^0\mb Q_{kk}\rb\propto ||Q_{kk}||_1$. 
For $n=0$ the trace of the Frobenius-Perron operator is not defined. The integral on the lhs of (\ref{trace}), however, 
is defined and gives the leading order contribution $N$. 
We replace the traces by sums of the Ruelle-Pollicott resonances (\ref{trres}) 
and make use of the symmetry ${\rm tr}{\cal P}^{-n}={\rm tr}{\cal P}^{n}$, 
\begin{equation}
\overline{||Q_{ik}||^2}^{\Delta\omega}(\omega)
=2\frac{N-1}{N}+\frac{4}{N}\sum_{n=1}^{\cal N}
\sum_\nu \lambda^n_\nu\cos n\omega 
\label{sumres}
\,.
\end{equation}
Note that the eigenvalue $1$ is also dropped in the leading order term.

To get from the smoothed $L^2$ norms $\overline{||Q_{ik}||^2}^{\Delta\omega}(\omega)$ 
to a mean value $\left\langle||Q_{ik}||^2\right\rangle_{\Delta\omega}(\omega)$ we have to 
divide it by the smoothed level density of the Husimi spectrum. 
The density of the Husimi spectrum is identified as 
the density-density correlation function with respect to the Floquet eigenphases, 
\begin{equation}
\rho_H(\omega)=\sum_{ik}\delta(\omega-(\phi_i-\phi_k))
\,.
\end{equation}
This spectral density can be calculated from the time Fourier transformation of 
the form factor. The smoothed density is given by a truncated Fourier transformation as 
\begin{equation}
\overline{\rho}^{\Delta\omega}(\omega)=\frac{N^2}{2\pi}\left(1+
\frac{2}{N^2}\sum_{n=1}^{\cal N} |{\rm tr}F^n|^2\cos n\omega\right)
\,,
\end{equation}
where we have separated the leading order term ($n=0$). 
It is known that the form factor is small as the time $n$ is small 
($\left\langle|{\rm tr}F^n|^2\right\rangle=n$ or $\approx 2n$ for CUE or 
COE, respectively). Thus, the summation up to the Ehrenfest time 
which is much smaller than $N$ becomes negligible and the smoothed spectral 
density is nearly constant. 
Before we write down the final result, we consider the truncated Fourier transformation 
of the resonances. Since the moduli of the resonances are smaller $1$, the 
summations in (\ref{sumres}) converge quickly such that we can replace the Ehrenfest time 
${\cal N}$ by $\infty$,
\begin{equation}
 \left\langle||Q_{ik}||^2\right\rangle(\omega)
=\frac{4\pi}{N^2}\left(
\frac{N-1}{N}+\frac{2}{N}\sum_\nu\sum_{n=1}^{\infty}
\lambda_\nu^n\cos n\omega\right)
\,.
\end{equation}
The constant term coincides with the RMT result (\ref{skewl2coe}), 
where $\frac{N}{N+1}=\frac{N-1}{N}+{\cal O}(N^{-2})$. 
The resonances lead to an $\frac{1}{N}$ (alias $\hbar$) correction in form of 
overlapping Lorentz distributions (see (\ref{lorentz})). 
In the classical limit 
the resonance corrections vanish as $N$ goes to infinity. 
However, for finite dimension we see non-universal corrections which 
are related to the chaoticity of the system. 
If, for instance, the classical dynamics is 
strongly chaotic such that all correlations vanish after one iteration, the 
resonances are close to the origin and the averaged $L^2$ norms show 
no deviations from RMT result. 
 
Before we come to numerical results we should discuss some prelimaries. 
The kicked top is known to have a mixed phase space. 
Although elliptic islands of stable periodic orbits are much smaller than the 
Planck cell, bifurcations can be responsible for further localization phenomena 
which are sometimes called super scars \cite{keating2}. 
But in contrast to the resonances, bifurcations strongly influence spectral correlations. 
That means that peaks resulting from bifurcating orbits are higher for the smoothed $L^2$ norms 
than for the averaged $L^2$ norms. 
Thus, we are able to distinguish between localization phenomena of resonances and bifurcations. 
The smoothing is again done by a convolution with a sinc function (\ref{smooth}). 
From the contributions of the diagonal Husimi eigenfunctions we have 
neglegted the squared $L^1$ norms which correspond to the stationary eigenvalue. 

\begin{figure}
\begin{center}
\leavevmode
\epsfxsize=1\textwidth
\epsffile{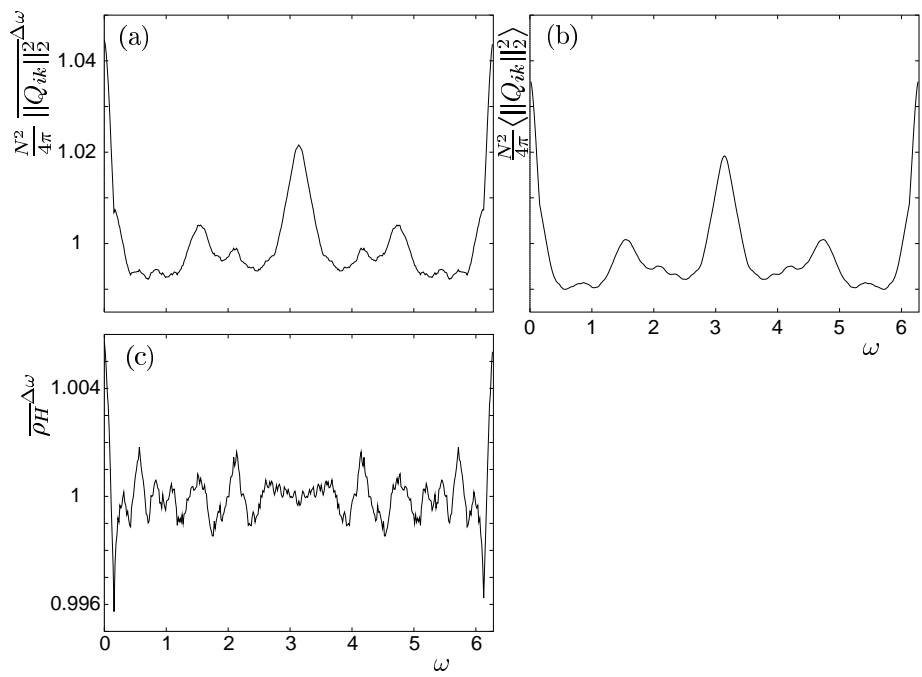}
\end{center}
\caption{(a) Smoothed and (b) averaged $L^2$ norms of skew Husimi eigenfunctions. 
(c) smoothed (Husimi) spectral density.}
\label{mitqd}
\end{figure}

\begin{figure}
\begin{center}
\leavevmode
\epsfxsize=1\textwidth
\epsffile{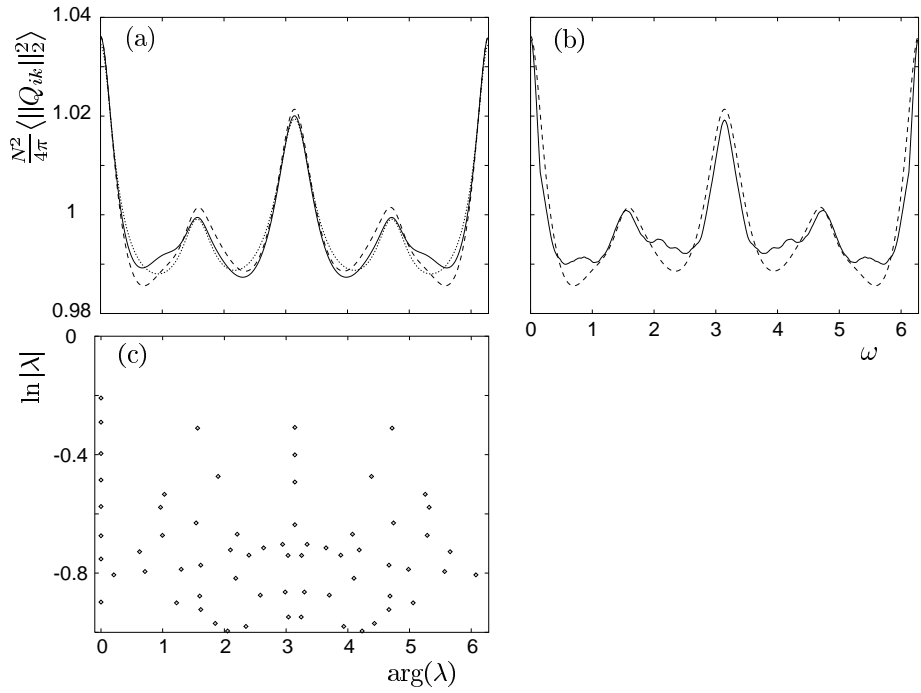}
\end{center}
\caption{ 
(a) Semiclassical predictions computed from 
(i) $8$ stabilized eigenvalues (dotted), 
(ii) eigenvalues with $|\lambda|>0.45$ (solid), 
and (iii) almost all eigenvalues of the truncated Frobenius-Perron matrix (dashed). 
The semiclassical prediction is mainly influenced from a few eigenvalues of large moduli. 
(b) Comparison of the semiclassical prediction (iii) (dashed) with averaged 
$L^2$ norms of skew Husimi eigenfunctions (solid). 
(c) Logarithmic plot of eigenvalues of the truncated Frobenius-Perron matrix 
($\ln|\lambda|$ versus $\arg\lambda$).
The peaks in (a) are associated to at least one eigenvalue of large modulus.}
\label{qmithsp}
\end{figure}

Fig.~\ref{mitqd} shows (a) the smoothed $L^2$ norms, (b) the mean $L^2$ norms, 
and (c) the smoothed spectral density. In (a) we see a couple of peaks 
at the Husimi eigenphases $\omega=0$, $\pi$, $\pm \frac{\pi}{2}$, and $\pm \frac{2\pi}{3}$. 
In (c) are remarkable peaks at the positions $\omega=\pm \frac{2\pi}{3}$ and after 
division we see in (b) that these peaks are suppressed, 
while the other peaks still have the same magnitude. 
Comparison with the semiclassical prediction is shown in Fig.~\ref{qmithsp}. 
Due to the fact that stabilized eigenvalues representing the resonances are 
of largest moduli we expect that the semiclassical prediction is almost independent 
of the set of eigenvalues as well as all stabilized ones are taken into account. 
In (a) the semiclassical prediction is computed from (i) eight stabilized eigenvalues, 
(ii) eigenvalues of modulus larger $0.45$, and (iii) almost all eigenvalues of the 
truncated Frobenius-Perron matrix. 
Comparison of (i)--(iii) shows that the semiclassical prediction is mainly influenced from 
a few classical eigenvalues of large moduli which represent the resonances. 
In (b) we compare the semiclassical 
prediction (iii) with the quantum result. 
In particular for the peaks we find a very good agreement. 
This result shows that the probability to find strongly overlapping eigenfunctions becomes 
large if the differences of their eigenphases coincides with the phase of a leading resonance, 
i.e. resonance of large modulus. 
In comparison with the background the peaks are small (a few percent). 
However, we show in the next section that strongly scarred eigenfunctions are 
mainly responsible for the peaks.  
The eigenvalues of the Frobenius-Perron matrix (c) 
are plotted logarithmically ($\ln |\lambda|$ versus $\arg\lambda$). 
These are easily associated with the peaks in (a). 
It should be remarked that there is no eigenvalue of large modulus which corresponds to the 
small peaks at $\omega=\pm\frac{2\pi}{3}$ in Fig.~\ref{mitqd} (a).

\section{Scarred Eigenstates of the Kicked Top}
\begin{figure}
\begin{center}
\leavevmode
\epsfxsize=1\textwidth
\epsffile{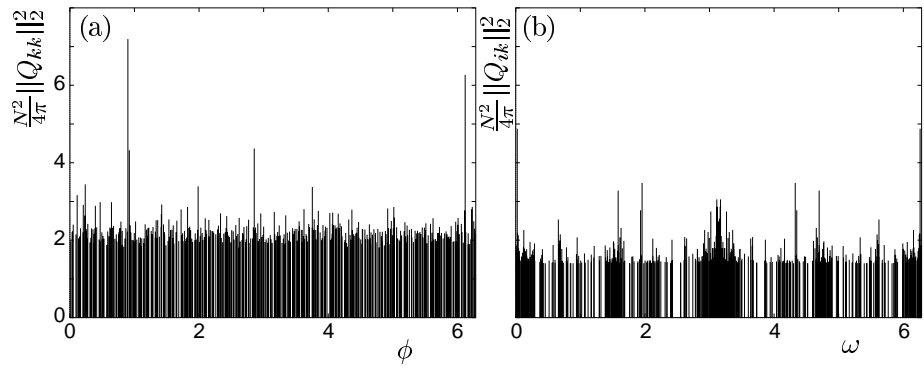}
\end{center}
\caption{(a) $L^2$ norms of diagonal Husimi eigenfunctions versus Floquet eigenphases. 
(b) $530$ $L^2$ norms of most strongly localized skew Husimi eigenfunctions. Note the frequent 
appearance close to the phases of resonances.}
\label{ql2dgn}
\end{figure}

In this section we consider single eigenfunctions and verify the statement that 
phase differences of strongly overlapping eigenfunctions coincide with 
phases of leading resonances. Due to the Schwarz inequality we first check 
that skew Husimi eigenfunctions with large $L^2$ norms are composed by scarred 
diagonal Husimi eigenfunctions. 
In Tab.~\ref{scartab} we find 25 $L^2$ norms of most strongly localized 
skew Husimi eigenfunctions, 
their eigenphases, and the corresponding diagonal Husimi eigenfunctions 
(the numbers correspond to the enumeration in Tab.~\ref{hustab}). 
Due to the symmetry $Q_{ik}=Q_{ki}^\ast$ we have restricted the eigenphases as $0<\omega\le\pi$. 
The 32 most strongly localized diagonal Husimi eigenfunctions are presented in Tab.~\ref{hustab}. 
Comparison of both tables proves that all skew Husimi eigenfunctions considered 
are composed from at least one scarred diagonal eigenfunction. 
In Fig.~\ref{ql2dgn} (a) $L^2$ norms of all diagonal Husimi eigenfunctions are shown, 
while in 
(b) we see 530 (of total $1.6\times10^{5}$) 
$L^2$ norms of most strongly localized skew Husimi eigenfunctions. 
Due to our foregoing results 
localized skew Husimi eigenfunctions appear frequently around the resonance phases. 
Interestingly, there are only two remarkable eigenfunctions (No 11 and 12 in Tab.~\ref{scartab})
around the Husimi eigenphases 
$\omega=\pm\frac{2\pi}{3}$. Moreover, in both cases one of the underlying diagonal 
Husimi eigenfunctions is 
No 16 in Tab.~\ref{hustab}. 
Further investigation of this eigenfunction has shown that it is strongly scarred on 
two bifurcating orbits of periods $1$ and $3$. 
It seems that we found a super scar corresponding to 
a period-tripling bifurcation.

\section{Conclusion}
In conclusion, phase-space localization of quantum \mbox{(quasi-)}eigenenergy 
functions, say scarring, is not only explained by periodic orbits, but also by 
Ruelle-Pollicott resonances and their corresponding resonance eigenfunctions. 
In particular, we found the interesting result that quantum Floquet eigenfunctions are pairwise 
localized in the same phase-space regions if the difference of their \mbox{(quasi-)}eigenenergies 
coincides with the phase of a leading resonance, i.e. resonance close to the unit circle. 
But note that this is a statistical statement which does not make a prediction for individual 
eigenstates. 
Moreover, we can not determine if there are either a few strongly scarred 
or many weakly localized eigenfunctions. 
However, the semiclassical prediction of the averaged $L^2$ norms is in a good agreement 
with numerical results. 

The correspondence between scars around periodic orbits described by Heller and 
the results of this paper might be 
understood as follows: resonances can be computed by a so-called cycle expansion, 
where resonances appear as roots of a polynomial whose coefficients are 
calculated from contributions of short periodic orbits (pseudo orbits) \cite{christiansen,cvit2}. 
On the one hand, scars typically appear around weakly unstable periodic orbits. 
On the other hand, these weakly unstable orbits mainly induce the cycle expansion. 

Although the kicked top has a mixed phase space, localization effects of stable orbits 
or bifurcations can be neglected if such phase-space structures are not resolved by the Planck cell. 
Further investigations are needed for the understanding 
of the so-called super scars which are related to bifurcations. 

For fruitful discussions the author thanks 
Fritz Haake, Karol {\.Z}yczkowski, 
Sven Gnutzmann, Joachim Weber, Petr Braun and Pierre Gaspard. 

We gratefully acknowledge support by the Sonderforschungsbereich 
Unordnung und gro{\ss}e Fluktuationen of the Deutsche 
Forschungsgemeinschaft.

\newpage

\begin{table}
\[
\begin{array}{|r|r|r||r|r|} \hline
{\rm No} & {\rm Re}\,\lambda & {\rm Im}\,\lambda 
& {\rm Re}\lambda_\nu &{\rm Im}\lambda_\nu \\ \hline
1 & 0.811497 & 0 & 0.81  & 0 \\ \hline
2 & 0.748029 & 0 & 0.75  & 0 \\ \hline
3 &-0.734887 & 0 &-0.745 & 0 \\ \hline
4 & 0.003495 &-0.733015 & 0 &-0.75 \\ \hline
5 & 0.003495 & 0.733015 & 0 & 0.75 \\ \hline
6 & 0.672874 & 0 & 0.67 & 0 \\ \hline
7 &-0.669930 & 0 &-0.68 & 0 \\ \hline
8 &-0.611201 & 0 &-0.63 & 0 \\ \hline
\end{array}
\]
\caption{Eigenvalues of the truncated Frobenius-Perron matrix 
(left column) which 
represent the expected resonance positions (right column). }
\label{restab}
\end{table}

\begin{table}
\[
\begin{array}{|r|r|r|r|r|} \hline
{\rm No} & \big[\phi_i-\phi_k\big] &\frac{N^2}{4\pi} ||Q_{ik}||^2
&{\rm No}&{\rm No} \\ \hline
1  &0.02293 &4.87904 &9  &8 \\ \hline
2  &0.02995 &2.17252 &4  &2 \\ \hline
3  &0.03690 &2.12795 &   &8 \\ \hline
4  &0.12504 &2.26266 &4  &1 \\ \hline
5  &0.66008 &2.53247 &8  &4 \\ \hline
6  &0.69002 &2.14943 &8  &2 \\ \hline
7  &0.95140 &2.07455 &32 &8 \\ \hline
8  &1.58566 &3.28086 &30 &11\\ \hline
9  &1.61324 &2.07295 &30 &  \\ \hline
10 &1.63071 &2.31385 &30 &  \\ \hline
11 &1.93314 &2.77164 &16 &9 \\ \hline
12 &1.95607 &3.47930 &16 &8 \\ \hline
13 &2.12224 &2.03752 &16 &  \\ \hline
\end{array}
\quad
\begin{array}{|r|r|r|r|r|} \hline
{\rm No} & \big[\phi_i-\phi_k\big] &\frac{N^2}{4\pi} ||Q_{ik}||^2
&{\rm No}&{\rm No} \\ \hline
14 &2.41853 &2.01226 &21 &  \\ \hline
15 &2.61615 &2.08038 &16 &4 \\ \hline
16 &2.64610 &2.04506 &16 &2 \\ \hline
17 &2.92433 &2.26189 &21 &5 \\ \hline
18 &2.97684 &2.21163 &   &8 \\ \hline
19 &2.99690 &2.05725 &   &9 \\ \hline
20 &3.01983 &2.73944 &   &8 \\ \hline
21 &3.10287 &2.10537 &23 &9 \\ \hline
22 &3.11230 &3.05349 &24 &8 \\ \hline
23 &3.11857 &2.86021 &17 &30\\ \hline
24 &3.12580 &2.55565 &23 &8 \\ \hline
25 &3.13523 &2.48017 &24 &9 \\ \hline
\end{array}
\]
\caption{Husimi eigenphases, $L^2$ norms of most strongly localized skew Husimi eigenfunctions, 
and the corresponding diagonal 
Husimi eigenfunctions (enumeration in Table~\ref{hustab}). 
Due to the rescaling the RMT-average is nearly $1$. 
}
\label{scartab}
\end{table}

\begin{table}
\[
\begin{array}{|r|r|r|} \hline
{\rm No} & \phi_k &\frac{N^2}{4\pi} ||Q_{kk}||^2 \\ \hline
1 & 0.11280& 3.16342\\ \hline
2 & 0.20789& 2.91329\\ \hline
3 & 0.23200& 2.63425\\ \hline
4 & 0.23784& 3.44206\\ \hline
5 & 0.39507& 2.88719\\ \hline
6 & 0.46842& 2.98415\\ \hline
7 & 0.64437& 2.98154\\ \hline
8 & 0.89792& 7.19529\\ \hline
9 & 0.92085& 4.32135\\ \hline
10& 1.41602& 2.67399\\ \hline
11& 1.42231& 2.92272\\ \hline
12& 1.72257& 2.79565\\ \hline
13& 1.82172& 2.86009\\ \hline
14& 1.98572& 3.39027\\ \hline
15& 2.33226& 2.69246\\ \hline
16& 2.85399& 4.36732\\ \hline
\end{array}
\quad
\begin{array}{|r|r|r|} \hline
{\rm No} & \phi_k & \frac{N^2}{4\pi}||Q_{kk}||^2 \\ \hline
17& 2.95522& 2.69166\\ \hline
18& 3.16445& 2.73181\\ \hline
19& 3.34145& 2.64167\\ \hline
20& 3.62419& 2.80308\\ \hline
21& 3.75393& 3.37501\\ \hline
22& 3.84585& 2.75931\\ \hline
23& 4.02372& 2.70190\\ \hline
24& 4.06881& 2.71314\\ \hline
25& 4.15409& 2.68584\\ \hline
26& 4.36618& 2.79003\\ \hline
27& 4.91979& 2.81994\\ \hline
28& 5.01485& 2.85668\\ \hline
29& 6.11281& 2.76855\\ \hline
30& 6.11984& 6.27044\\ \hline
31& 6.21927& 2.79705\\ \hline
32& 6.22970& 2.86170\\ \hline
\end{array}
\]
\caption{Floquet eigenphases and $L^2$ norms of most strongly 
localized diagonal Husimi eigenfunctions. 
Note that the RMT-average is about $2.034$}
\label{hustab}
\end{table}

\end{document}